\documentstyle[12pt,preprint,aps,epsfig]{revtex}

\begin{document}
\draft

\title{   Infinite Order Discrete Variable
Representation for Quantum Scattering}

\author{Nark Nyul Choi, Min-Ho Lee} 
\address{Department of Physics, Kumoh National University of Technology,
Kumi 730-701, Korea}
\author{Sung Ho Suck Salk} 
\address{Department of Physics, Pohang University 
	 of Science and Technology, Pohang 790-600, Korea}
\date{\today}
\maketitle

\begin{abstract}
A new approach to multi-dimensional quantum scattering by the 
infinite order discrete variable representation is presented.
Determining the expansion coefficients of the wave function 
at the asymptotic regions by the solution of the differential
Schr\"{o}dinger equation, 
we reduce an infinite set of linear equations to
a finite one. Application to the benchmark collinear H + H$_2$
$\rightarrow $ H$_2$ + H reaction is shown to yield 
precise reaction probabilities.
\end{abstract}
\vskip1cm
\pacs{PACS numbers: 34.10.+x, 34.50.-s, 34.50.Pi, 03.65.Nk, 03.80.+r}

One of the most common approach to the solution of quantum scattering
problem is the application of square-integrable ($L^2$) basis 
functions\cite{Rescigno,Heller,Zhang,Schwenke,Aymar}.
Of the $L^2$ basis methods 
the discrete variable representation(DVR) method \cite{Harris} 
is proven to be highly successful \cite{Light,Colbert,Thompson}.
The DVR is a grid-point representation in which the potential 
energy is diagonal and the kinetic energy is a sum of one-dimensional
matrices. Hence the  Hamiltonian matrix is extremely sparse, which 
means that iterative linear algebra methods \cite{qmr} 
can deal efficiently with extremely large systems.

Recently, Eisenberg {\it et. al.} \cite{baer} 
has developed an infinite order
DVR method for one-dimensional quantum reactive scattering problems.
They expanded the wave function in terms of an infinite set of 
$L^2$ basis functions satisfying the conditions of 
DVR \cite{Harris,Light,Colbert,muckerman}.
The matrix related to the resulting set of algebraic equations,
though infinite, has been shown to have the structure of a 
Toeplitz matrix\cite{toeplitz}.
Using the  analytical properties  of the Toeplitz matrix,
they reduced the infinite set of algebraic equations to a 
finite one and obtained very accurate results for one-dimensional
potential scattering.
However, their method is not likely to be extended to
multi-dimensional systems 
due to the failure of the Toeplitz properties.

In this Letter, we show that the infinite order DVR can 
be properly converted into a finite DVR even for multi-dimensional
reactive systems.  Instead of investigating the analytical properties
of the Toeplitz matrix, we use the solutions of the differential
Schr\"{o}dinger equations at the asymptotic regions.

The full scattering wave function $\Psi_n$ is decomposed of 
the incoming distorted wave
$\theta^-_n$
and the outgoing wave $\chi^+_n$,
\begin{equation}
\Psi_n = -\theta^-_n + \chi^+_n,
          \label{fullwave}
\end{equation}
where $n$ is a channel index, i.e., a superindex over the arrangement 
and  rovibrational indices. The distorted wave
$\theta^-_n$ is any regular scattering
solution corresponding to a simple nonreactive 
(i.e., arrangement conserving)
Hamiltonian, $H_{0n}$, 
in the asymptotic region in the arrangement channel $n$:
\begin{equation}
( H_{0n} - E) \theta^-_n = 0 ,
\end{equation}
while $\Psi_n$ is the solution of
\begin{equation}
  ( H - E) \Psi_n = 0,
\end{equation}
where $H$ is the full Hamiltonian.
For the sake of later convenience, we impose
the "totally incoming"  boundary condition on $\theta^-_n$ as
\begin{equation}
\theta^-_n (r,R)  \sim  \frac{ e^{- i k_n R}}{\sqrt{v_n} }
		 u_n(r)
\end{equation}
for large R.
Here $r$ denotes all possible internal coordinates of the system,
thus excluding the channel radius (relative translation)
$R$; $\{u_n(r)\}$, the channel eigenfunctions;
$v_n$ the relative velocity for channel $n$, and 
$k_n$ the corresponding wave vector. The equation for $\chi^+_n$ is 
then, from (1)-(3),
\begin{equation}
(E - H) \chi^+_n = (H - E)\theta^-_n.
\end{equation}
$\chi^+_n$ must obey "totally outgoing" boundary condition,
i.e., the only incoming wave part in the full wave function is due to 
$\theta^-_n$.

Using the infinite order uniform DVR \cite{Colbert}, 
(5) is transformed into 
an infinite set of coupled linear algebraic equations.
To do this, we introduce the following convenient sets 
of DVR basis functions
${\cal Q},{\cal R}_0,{\cal P}_0,R$ 
and ${\cal P}$ as shown in Fig. 1:
${\cal Q}$ be the set of $N$ DVR basis functions 
represented by the $N$ grid points 
in the reactive (strong) interaction region correspond;
${\cal R}_0 = \{ {\cal R}_0(1),...,{\cal R}_0(N_r) \}$,
the set of $N_r$ DVR basis functions
in the reactant asymptotic (relatively weak interaction) region
nearest to the region of reactive interaction;
${\cal P}_0 = \{ {\cal P}_0(1),...,{\cal P}_0(N_p) \}$,
the set of $N_p$ DVR basis functions
in the product asymptotic regions nearest to the
region of reactive interaction;
${\cal R}$, the infinite set of DVR basis functions
in the reactant asymptotic region except the functions
in ${\cal R}_0$, and ${\cal P}$ the infinite set of DVR basis functions
in the product asymptotic regions 
except the functions in ${\cal P}_0$.
Here $N_r(N_p)$ is the number of open channels in the reactant (product)
arrangement.
Using these sets, we can rewrite (5)
as a set of coupled algebraic equations,
\begin{eqnarray}
A_{{\cal R}_0{\cal R}} \langle {\cal R}|\chi^+_{n_r} \rangle  +
A_{{\cal R}_0{\cal R}_0} \langle {\cal R}_0|\chi^+_{n_r} \rangle  +
A_{{\cal R}_0{\cal Q}} \langle {\cal Q}|\chi^+_{n_r} \rangle  +
A_{{\cal R}_0{\cal P}_0} \langle {\cal P}_0|\chi^+_{n_r} \rangle  +
A_{{\cal R}_0{\cal P}} \langle {\cal P}|\chi^+_{n_r} \rangle  
&=& \langle {\cal R}_0 |H-E|\theta^-_{n_r} \rangle ,   
    \\
A_{{\cal Q}{\cal R}} \langle {\cal R}|\chi^+_{n_r} \rangle  +
A_{{\cal Q}{\cal R}_0} \langle {\cal R}_0|\chi^+_{n_r} \rangle  +
A_{{\cal Q}{\cal Q}} \langle {\cal Q}|\chi^+_{n_r} \rangle  +
A_{{\cal Q}{\cal P}_0} \langle {\cal P}_0|\chi^+_{n_r} \rangle  +
A_{{\cal Q}{\cal P}} \langle {\cal P}|\chi^+_{n_r} \rangle  
&=& \langle {\cal Q} |H-E|\theta^-_{n_r} \rangle ,
    \\
A_{{\cal P}_0{\cal R}} \langle {\cal R}|\chi^+_{n_r} \rangle  +
A_{{\cal P}_0{\cal R}_0} \langle {\cal R}_0|\chi^+_{n_r} \rangle  +
A_{{\cal P}_0{\cal Q}} \langle {\cal Q}|\chi^+_{n_r} \rangle  +
A_{{\cal P}_0{\cal P}_0} \langle {\cal P}_0|\chi^+_{n_r} \rangle  +
A_{{\cal P}_0{\cal P}} \langle {\cal P}|\chi^+_{n_r} \rangle  
&=& \langle {\cal P}_0 |H-E|\theta^-_{n_r} \rangle  ,    
\end{eqnarray}
where 
\begin{equation}
A_{ij} = (E - V_j) \delta_{ij} - T_{ij},
\end{equation}
$T_{ij}$ is the kinetic energy matrix element\cite{Colbert},
which is analytically obtained,
connecting the DVR grid points 
$i \in {\cal R}_0 + {\cal Q} + {\cal P}_0$ 
and
$j \in {\cal R} + {\cal R}_0 + {\cal Q} + {\cal P}_0 + {\cal P}$,
$V_j$ is the potential energy at the DVR grid point $j$,
and we omit the summation over the index $j$ such that, e.g.,
$A_{{\cal R}_0{\cal R}} \langle {\cal R}|\chi^+_{n_r} \rangle$
means
$\sum_{j \in {\cal R} + {\cal R}_0 + {\cal Q} + {\cal P}_0 + {\cal P}}
A_{ij} \langle j | \chi^+_{n_r} \rangle$
for $i \in {\cal R}_0$.

In the above we did not write explicitly
the similar equations corresponding to
the ${\cal R}$ and ${\cal P}$ component 
since we do not use them in order to eliminate 
$\langle {\cal R}|\chi^+_{n_r} \rangle$ and 
$\langle {\cal P}|\chi^+_{n_r} \rangle$
by searching for an analytical property as in the case of Toeplitz.
Instead, for the asymptotic regions, 
we introduce,
\begin{eqnarray}
\langle {\cal P}|\chi^+_{n_r} \rangle &=& 
   - \sum_{n_p =1}^{N_p} \langle {\cal P}|\theta^+_{n_p} \rangle 
     S_{n_pn_r},       \\
\langle {\cal R}|\chi^+_{n_r} \rangle &=& 
   - \sum_{n_r' =1}^{N_r} \langle {\cal R}|\theta^+_{n_r'} \rangle 
     S_{n_r'n_r},      
\end{eqnarray}
where $\theta^+_n$ is the regular "totally outgoing" wave which 
satisfies (2), i.e., $\langle {\cal P}|\theta^+_n \rangle =
\langle {\cal P}|\theta^-_n\rangle^*$,
and $S$ is the scattering matrix to be determined.
(10) and (11) can be rewritten as 
\begin{eqnarray}
\langle {\cal P}|\chi^+_{n_r}\rangle &=&
\sum_{n_p=1}^{N_p} \sum_{i=1}^{N_p} \langle {\cal P} |
\theta^+_{n_p} \rangle
\left( T^{\cal P} \right)^{-1}_{n_p i} 
\langle {\cal P}_0(i) | \chi^+_{n_r}\rangle, 
\\
\langle {\cal R}|\chi^+_{n_r}\rangle &=&
\sum_{n_r'=1}^{N_r} \sum_{i=1}^{N_r}
\langle {\cal R} | \theta^+_{n_r'} \rangle
\left( T^{\cal R} \right)^{-1}_{n_r' i} 
\langle {\cal R}_0(i) | \chi^+_{n_r}\rangle, 
\end{eqnarray}
where
\begin{equation}
T^{\cal P}_{i n_p} \equiv \langle {\cal P}_0(i) |\theta^+_{n_p}
    \rangle
\end{equation}
and
\begin{equation}
T^{\cal R}_{i n_r} \equiv \langle {\cal R}_0(i) |\theta^+_{n_r} \rangle.
\end{equation}
{\it The above equations} (12)-(15) 
{\it are the key in this Letter}.
Substituting $\langle {\cal P}|\chi^+_{n_r} \rangle $ and 
$\langle {\cal R}|\chi^+_{n_r} \rangle$ in (12) and (13) 
into (6) through (8), we obtain 
the following set of $N+N_r+N_p$ linear equations:
\begin{eqnarray}
(\epsilon + A)_{{\cal R}_0{\cal R}_0} \langle {\cal R}_0|\chi^+_{n_r}
\rangle +
 A_{{\cal R}_0{\cal Q}} \langle {\cal Q}|\chi^+_{n_r} \rangle +
(\epsilon + A)_{{\cal R}_0{\cal P}_0} 
\langle {\cal P}_0|\chi^+_{n_r} \rangle 
&=& \langle {\cal R}_0|H - E|\theta^-_{n_r}\rangle,
\\
(\epsilon + A)_{{\cal Q}{\cal R}_0} \langle {\cal R}_0|\chi^+_{n_r} 
\rangle +
 A_{{\cal Q}{\cal Q}} \langle {\cal Q}|\chi^+_{n_r} \rangle +
(\epsilon + A)_{{\cal Q}{\cal P}_0} \langle {\cal P}_0|\chi^+_{n_r} 
\rangle 
&=& \langle {\cal Q}|H - E|\theta^-_{n_r}\rangle,
\\
(\epsilon + A)_{{\cal P}_0{\cal R}_0} \langle {\cal R}_0|\chi^+_{n_r} 
\rangle +
 A_{{\cal P}_0{\cal Q}} \langle {\cal Q}|\chi^+_{n_r} \rangle +
(\epsilon + A)_{{\cal P}_0{\cal P}_0} \langle {\cal P}_0|\chi^+_{n_r} 
\rangle 
&=& \langle {\cal P}_0|H - E|\theta^-_{n_r}\rangle,
\end{eqnarray}
where $\epsilon$ is the matrix 
of which the non-vanishing elements are
\begin{equation}
\epsilon_{i{\cal R}_0} =
    \sum_{j \in {\cal R}}\sum_{n_r'} A_{ij} 
    \langle j | \theta^+_{n_r'}\rangle
    \left( T^{\cal R} \right)^{-1}_{n_r'{\cal R}_0}
\end{equation}
and
\begin{equation}
\epsilon_{i{\cal P}_0} =
    \sum_{j \in {\cal P}}\sum_{n_p} A_{ij} 
    \langle j | \theta^+_{n_p}\rangle
    \left( T^{\cal P} \right)^{-1}_{n_p{\cal P}_0},
\end{equation}
where $i$ is a DVR grid point in the sets
${\cal R}_0$, ${\cal Q}$, and ${\cal P}_0$.
Note that $\epsilon_{ij}$ is nonvanishing only for $j$
in the sets ${\cal R}_0$ and ${\cal P}_0$.
Applying the complex conjugate of (12)-(15), (19) and (20), 
the right hand sides of expressions (16) through (18)
can be expressed as
\begin{equation}
\langle i | H - E | \theta^-_{n_r} \rangle =
  - \left( A + \epsilon^* \right)_{i {\cal R}_0}
    \langle {\cal R}_0 | \theta^-_{n_r} \rangle .
\end{equation}
Hence, (16)-(18) can be expressed 
in the form of an $(N+N_r+N_p) \times (N+N_r+N_p)$ matrix equation
\begin{equation}
{\tilde{\bf A}} \cdot {\bbox {\chi}}^+_{n_r} =
   - {\tilde{\bf A}}^* \cdot {\bbox {\theta}}^-_{n_r},
\end{equation}
where 
${\tilde{\bf A}} = {\bf A} + {\bbox {\epsilon}}$.
Let ${\bf G}^+$ be the left-inverse of 
${\tilde{\bf A}}$,
i.e., ${\bf G}^+ \cdot {\tilde{\bf A}} = 1$. 
Then it can be shown that
\begin{equation}
{\bbox {\chi}}^+_{n_r} = 
    - {\bbox {\theta}}^-_{n_r} 
    + {\bf G}^+ \cdot ( {\bbox {\epsilon}} - {\bbox {\epsilon}}^*) 
    \cdot
    {\bbox  {\theta}}^-_{n_r}.
\end{equation}
Finally we obtain the full scattering
wave function and the scattering matrix,
\begin{equation}
{\bf \Psi}_{n_r} = 
{\bf G}^+ \cdot ({\bbox {\epsilon}} - {\bbox {\epsilon}}^*)
    \cdot {\bbox {\theta}}^-_{n_r},
\end{equation}
and
\begin{equation}
S_{n_pn_r} = 
   \sum_{i \in {{\cal P}_0} ,j \in {{\cal R}_0}} 
   \left( {\bf T}^{\cal P} \right)^{-1}_{n_p i}
   \left[ {\bf G}^+ \cdot 
   \left( {\bbox {\epsilon}}^* - {\bbox {\epsilon}}
			\right) \right]_{ij}
   \left( {\bbox {\theta}}^-_{n_r} \right)_j.
\end{equation}
For the calculation of ${\bbox {\epsilon}}$, we need the overlaps
$\langle {\cal R}|\theta^+_{n_r'} \rangle$,
$\langle {\cal R}_0 |\theta^+_{n_r'} \rangle$,
$\langle {\cal P}|\theta^+_{n_p} \rangle$, 
and 
$\langle {\cal P}_0 |\theta^+_{n_p} \rangle$. 
Using the property of the DVR basis functions,
\begin{equation}
\langle i | \theta^+_n \rangle =
\theta^+_n \left( r_i , R_i \right) \sqrt{w} ,
\end{equation}
where
$r_i$ and $R_i$ are respectively
the values of $r$ and $R$ at the DVR grid point $i$,
and 
$w$ is the quadrature weight \cite{Harris,Light,Colbert,muckerman},
the above overlaps can be easily
obtained by numerical integration of the
differential Sch\"{o}dinger equation (2)
corresponding to the nonreactive scattering region
discussed earlier.

Thus far we have shown that the derived scattering matrix (25)
is completely general enough to deal with the fully three dimensional
scattering processes.
For the sake of comparison with other accurate calculations,
we apply the method to the collinear 
H + H$_2$ $\rightarrow $ H$_2$ + H reaction 
in the energy range of 0.4$-$1.0 eV.
The coordinate axes for the DVR are chosen to be the normal ones
of the transition state \cite{dvr-abc} as shown in Fig. 1.
We use the LSTH potential energy surface \cite{lsth}.
The distorted wave functions, $\theta^+_n$, are calculated 
by integrating the coupled differential equations (2)
inward from $R_{asymp}$($\ge$ 20.0 a.u.) to $R_{max}$($\ge$ 6.0 a.u.)
for each arrangement, and then are interpolated at the DVR grid points
in the asymptotic regions to obtain the overlaps
$\langle {\cal R} | \theta^+_{n_r'} \rangle$, 
$\langle {\cal R}_0 | \theta^+_{n_r'} \rangle$, etc.
$E_1$=0.78629 eV is the threshold of the vibrational quantum states
with
$n=1$. Hence there are two linearly independent distorted wave functions
$\theta^+_0$ and $\theta^+_1$ at the energies above $E_1$.

Fig. 2 illustrates the convergence of $P_0 (E)$ and $P_1 (E)$
at $E=0.9678$ eV,
where $P_{n_r} (E)$ is 
the total reaction probability
from a given reactant molecular state $n_r$ 
to any product molecular states at energy $E$,
i.e.,
\begin{equation}
P_{n_r} (E) = \sum_{n_p = 1}^{N_p}
	      | S_{n_p n_r} |^2 ,
\end{equation}
as the size of the region of reactive 
interaction, $R_{max}$, is increased, for different values
of the grid constant $n_B$ \cite{Colbert,dvr-abc}.
The percent error is calculated by comparison to the results
of Colbert and Miller \cite{Colbert}. 
The convergence for $E=0.9678$ with respect to $n_B$ 
for $R_{max}=8.0$ is also shown in Fig. 3.
The total reaction probabilities, with $R_{max}=8.0$ and $n_B=3.4$,
are converged to within 0.05$\%$ at this energy,.
Fig. 4 shows excellent agreement of 
the converged results for $P_0 (E)$ and $P_1 (E)$ 
with the exact results of Bondi and Connor \cite{Bondi}
over a range of total energies including the resonance energy
near 0.9 eV.

In summary, the infinite order DVR theory for
multi-dimensional reactive scattering has been presented.
By applying it to the
collinear H + H$_2$
$\rightarrow $ H$_2$ + H reaction,
we obtained precise reaction probabilities comparable with
the other results obtained from the direct integration of 
the differential 
Schr\"{o}dinger equation (3) \cite{Bondi}.
The advantages of the present theory are as follows:
1. iterative linear algebra methods are readily applicable,
2. no numerical integrals are needed
for calculating the matrix elements,
3. as can be seen from (19),(20),(24) and (25),
no regularization techniques \cite{Zhang,Colbert}
for the distorted wave functions
are necessary since
no informations on those functions in the
reactive (strong) interaction region are required
although they are set to be regular at $R=0$ 
in the beginning (1),
4. the effective Green's operator ${\bf G}^+$ 
in the reactive interaction region is obtained as a by-product,
5. the formalism is completely general and can be applied
to any non-collinear scattering problems
including collinear systems.

We would like to thank D.G. Choi for helpful discussions
and W.H. Thompson for providing information on
the calculational results 
in references \cite{Bondi} and \cite{dvr-abc}.
This work was partially supported by the Center for Molecular Science
at KAIST, and the numerical calculations were performed
on the computer CRAY-C90 at SERI.


\begin{figure}
\caption{
Schematic diagram showing the sets ${\cal R}$ (plus),
${\cal R}_0$ (diamond),
${\cal Q}$ (dot),
${\cal P}_0$ (square),
and 
${\cal P}$ (cross) 
of the DVR basis functions, which are represented by the DVR
grid points,
for the collinear
H + H$_2$
$\rightarrow $ H$_2$ + H reactive scattering problem.
Three straight lines are drawn to divide
the sets ${\cal R}$, ${\cal Q}$, and ${\cal P}$
from each others.
The potential energy contours and the $(x,y)-$coordinate system for
the DVR are also depicted.
}
\vskip1cm
\caption{
Relative percent error of the total reaction probabilities $P_0 (E)$ 
(solid lines) and $P_1 (E)$ (dashed lines)
at $E=0.9678$ eV
as a function of $R_{max}$ for different values of $n_B$;
$n_B = 3.4$ (thick lines); and $n_B = 1.7$ (thin lines).
}
\vskip1cm
\caption{
Relative percent error of the total reaction probabilities $P_0 (E)$ 
(solid lines) and $P_1 (E)$ (dashed lines)
at $E=0.9678$ eV
as a function of $n_B$ for $R_{max}=8.0$.
}
\vskip1cm
\caption{
Total reaction probabilities (a) $P_0 (E)$ and (b) $P_1 (E)$ 
as a function of energy
computed using
$R_{max} = 8.0$ a.u. and $n_B = 3.4$ (circle)
compared to the results of Bondi and Connor 
(solid line) (ref. [15])
and Thompson and Miller (triangle) (ref. [16]).
}
\end{figure}


\newpage
\begin{figure}[bh]
\vskip2cm
\centerline{
\epsfig{figure=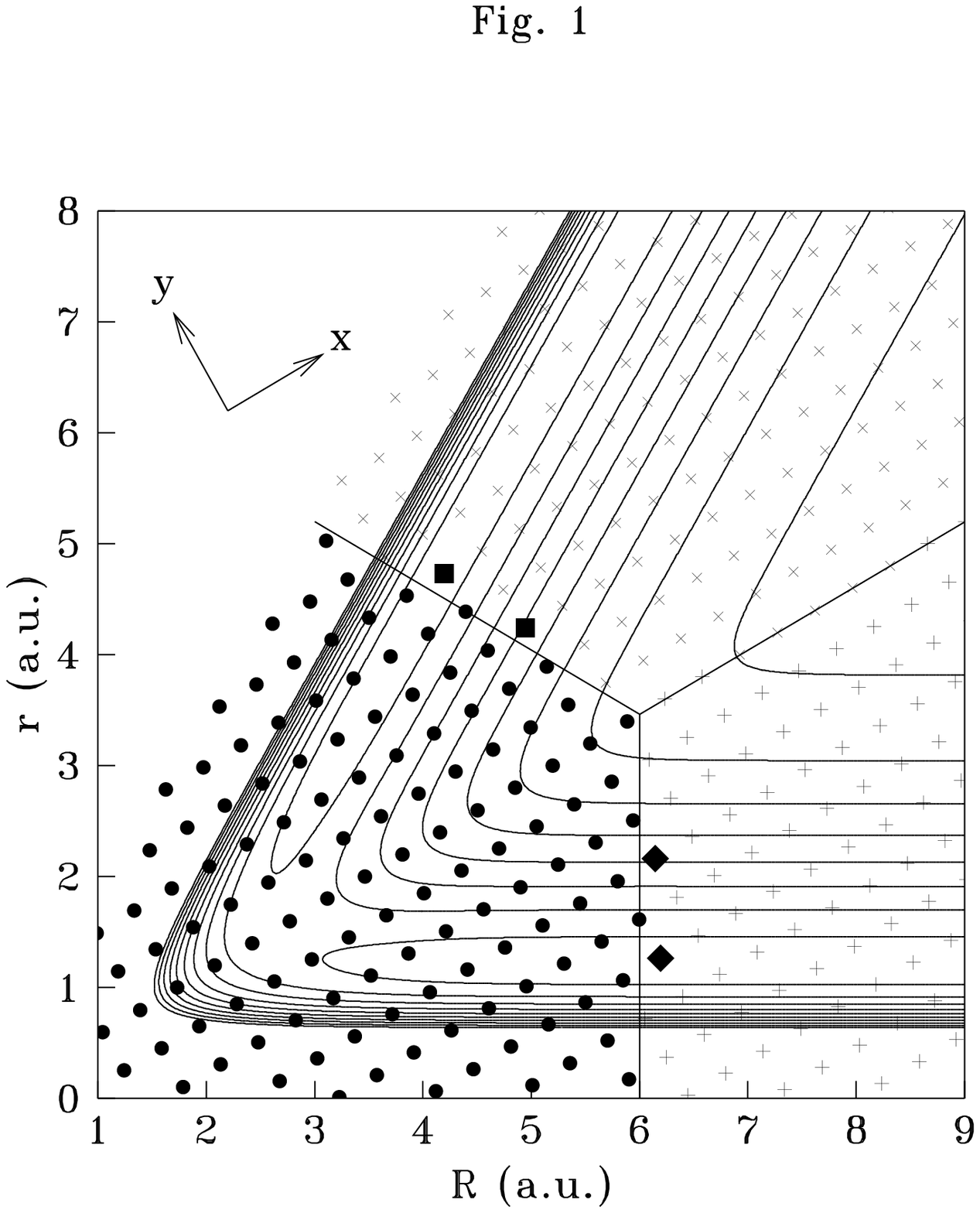,height=9cm}}
\end{figure}

\begin{figure}[bh]
\centerline{
\epsfig{figure=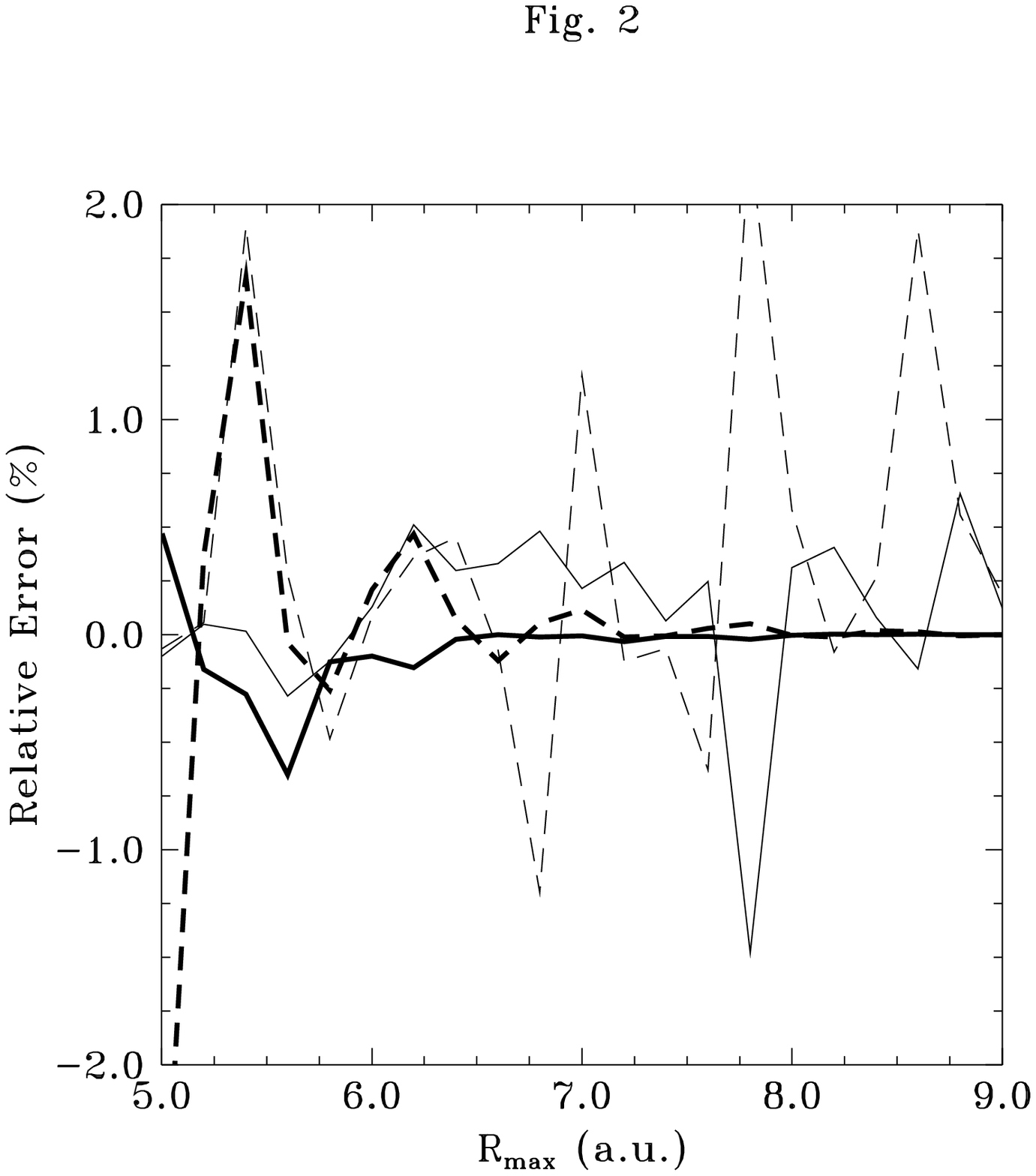,height=9cm}}
\end{figure}

\begin{figure}[bh]
\centerline{
\epsfig{figure=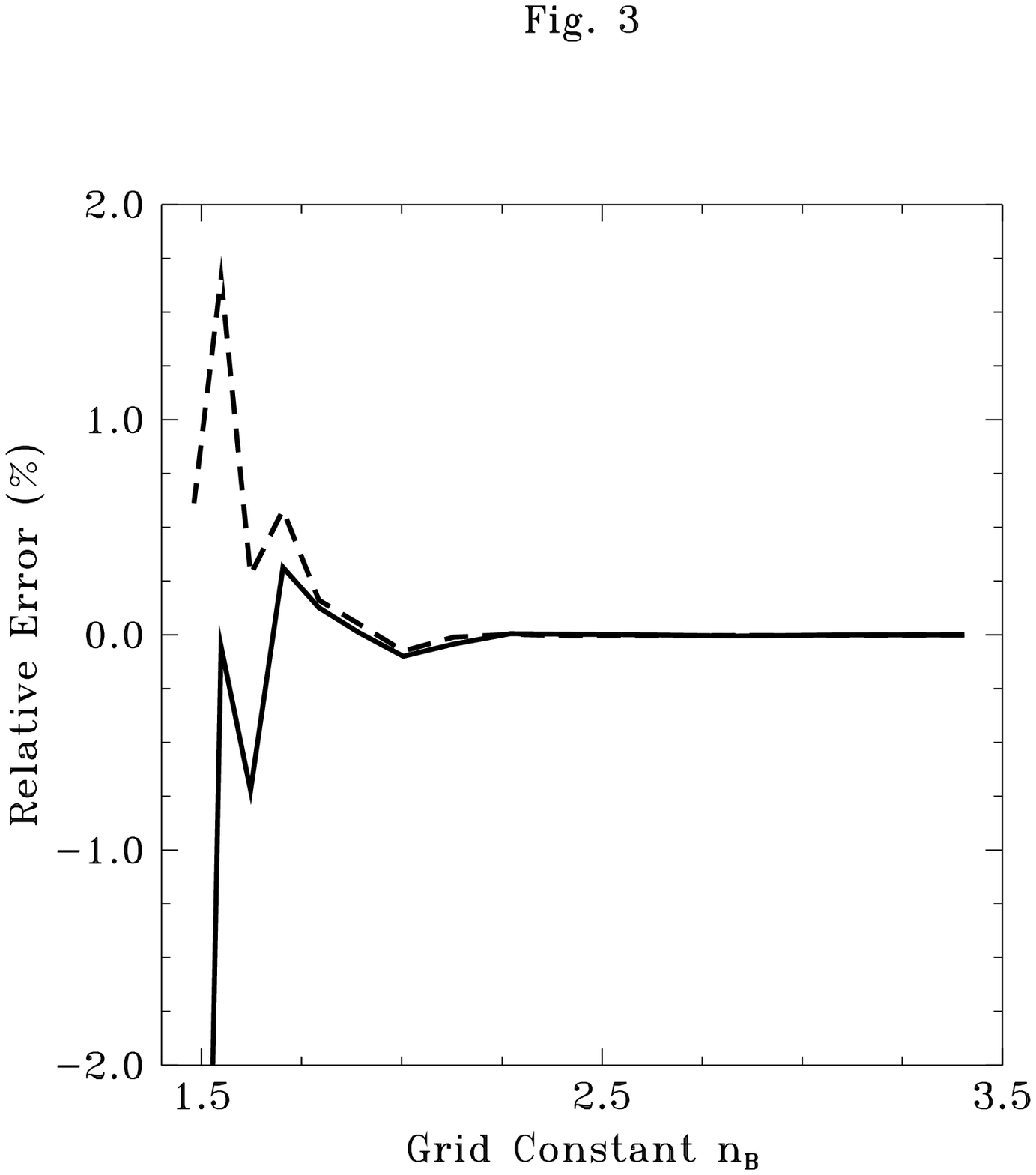,height=9cm}}
\end{figure}

\begin{figure}[bh]
\centerline{
\epsfig{figure=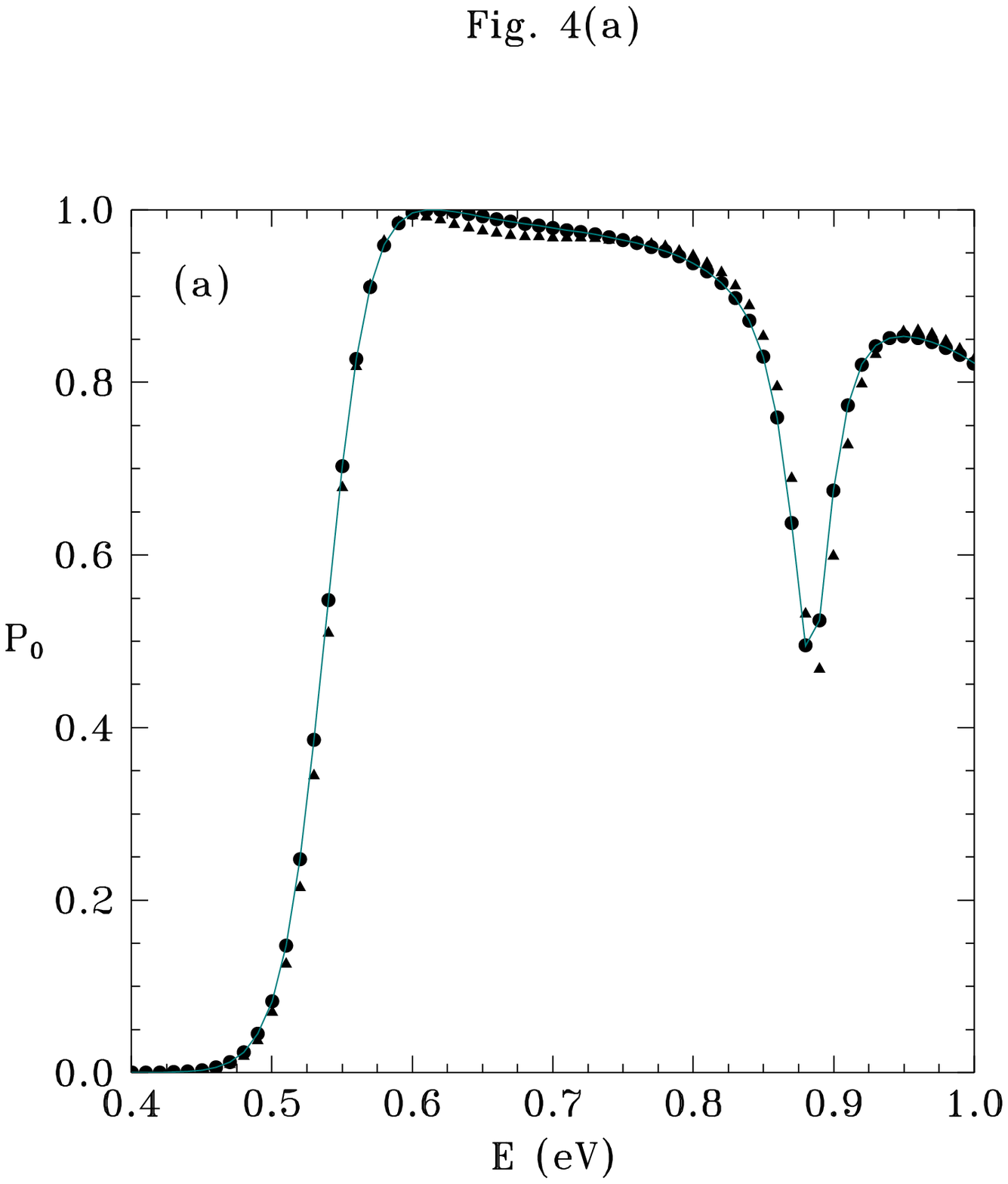,height=9cm}}
\end{figure}

\newpage
\begin{figure}[bh]
\centerline{
\epsfig{figure=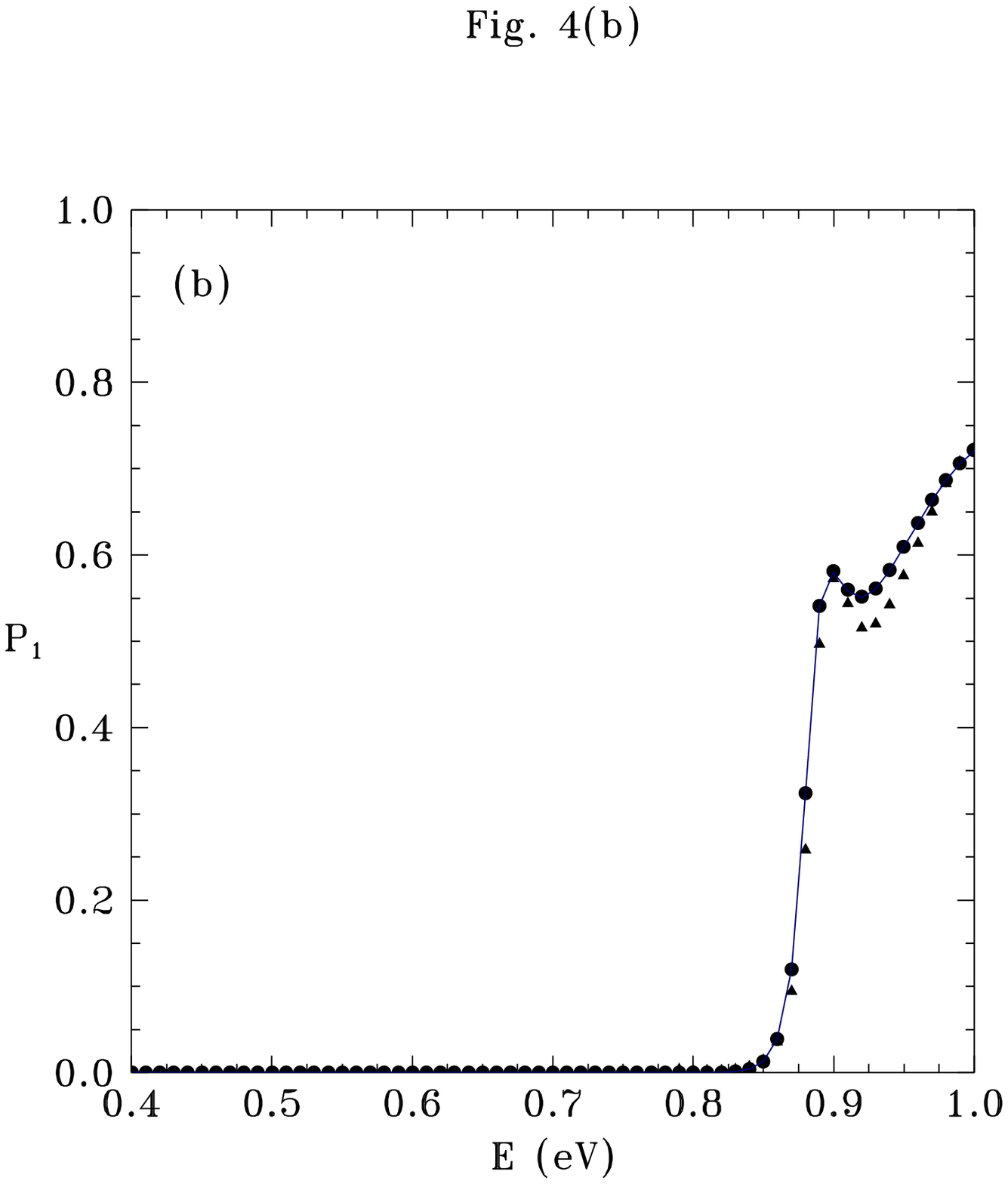,height=9cm}}
\end{figure}

\end{document}